\documentclass[11pt]{iopart}
\jl{33}
\usepackage{iopams,multirow,graphicx}
\begin{document}

\title[Phonon distributions of a single bath mode coupled to a quantum dot]{Phonon distributions of a single bath mode coupled to a quantum dot}

\author{F. Cavaliere\dag, G. Piovano\dag, E. Paladino\ddag, M. Sassetti\dag}
\address{\dag Dipartimento di Fisica, Universit\`a di Genova, LAMIA CNR-INFM, Via Dodecaneso 33, 16146 Genova, Italy}
\address{\ddag MATIS CNR-INFM, Catania \& Dipartimento di Metodologie Fisiche e Chimiche, Universit\`a di Catania, 95125 Catania, Italy}

\begin{abstract}
  The properties of an unconventional, single mode phonon bath coupled
  to a quantum dot, are investigated within the rotating wave
  approximation. The electron current through the dot induces an out
  of equilibrium bath, with a phonon distribution qualitatively
  different from the thermal one. In selected transport regimes, such
  a distribution is characterized by a peculiar selective population
  of few phonon modes and can exhibit a sub-Poissonian behavior. It is
  shown that such a sub--Poissonian behavior is favored by a double
  occupancy of the dot. The crossover from a unequilibrated to a
  conventional thermal bath is explored, and the limitations of the rotating wave
  approximation are discussed.
\end{abstract}

\pacs{73.23.-b;03.65.Yz} 
\submitto{\NJP -- Focus on Quantum Dissipation in Unconventional Environments}
\maketitle

\section{Introduction}
The study of dissipation and decoherence is a central problem in the
description of solid state systems both from the fundamental and the
applicative point of view. One of the most popular models, employed to
describe dissipation, is the one introduced by Caldeira and
Leggett~\cite{caldeira} in which the dissipative environment is a bath
of harmonic oscillators whose fluctuations obey Gaussian statistics,
linearly coupled to the system under
consideration~\cite{leggett,weiss}.  While this model is reasonable to
describe the dissipation induced by large environments, such as
equilibrium Fermi reservoirs, it is becoming clear that possible
sources of decoherence do not necessarily fall into this category for
nanodevices such as quantum dots and qubits.  Solid state nanocircuits
typically suffer from discrete noise, often originated from background
charge fluctuations. Single/few impurities behaving either like
random telegraph noise sources~\cite{duty} or being entangled with the
device have been recently observed~\cite{simmonds-ustinov} in
different setups. The resulting non-gaussian effects can only be
predicted introducing non-linear bath models~\cite{models}.  In
addition, the excellent control recently achieved in circuit-QED
experiments~\cite{walraff} has paved the way to the observation of
quantum effects originating from the coupling of multilevel
nanodevices to a single harmonic mode of a high-Q cavity resulting in
a effective bath having a structured power spectrum~\cite{NJP07-NPG}.
All these mechanisms deviate from the canonical Caldeira and Leggett
description and are examples of unconventional dissipative baths.

The coupling to an unconventional dissipative bath arises also in the
case of a nanoresonator coupled to a quantum dot, where electrons are
coupled to few or even a single phonon mode~\cite{debald}. An experimental realization of such a system
is the phonon cavity studied by Weig {\em et al.}~\cite{weig}. Here,
the coupling to the single bath mode induces sizable effects on the
transport properties of the dot, with the appearance of peaks in the
conductance spectrum when the energy of tunneling electrons matches
the energy of the phonon mode. A similar phenomenon occurs in a
oscillating C$_{60}$ molecule contacted by gold leads~\cite{park}.

Often, in these cases both the leads bath and the localized bath mode
are assumed to be in their thermal equilibrium~\cite{braig}. However,
there are several situations in which this assumption is not justified
especially for the localized mode. For example, in suspended carbon
nanotubes, the electrons are coupled with a single longitudinal
stretching~\cite{sapmaz}, or to a radial breathing~\cite{leroy} phonon
mode. The {\em unconventional} single mode bath is driven out of
equilibrium by the electron current and it induces the appearance of
peculiar negative differential conductance
traces~\cite{boese,zazunov}. Another example is that of the
superconducting single electron transistor coupled to a single--mode
nanoresonator studied by Naik {\em et al.}~\cite{naik}, where the
back--action exerted by the single electron transistor cools down the
phonon mode below the temperature of the environment. Then, it may be
necessary to treat the leads and the single mode on a different
footing. While the leads degrees of freedom can be assumed in their
thermal equilibrium and are traced out in the usual way, the coupled
dynamics of the electrons and the localized boson is treated without
assuming {\em a priori} a distribution for the localized mode.

Such a issue will be the subject of investigation in this work.
We will consider a spin degenerate single
level quantum dot with finite Coulomb repulsion coupled to two external environments:

\noindent ($i$) a pair of Fermi leads, described by a standard Caldeira Leggett model,  

\noindent ($ii$) an unconventional {\em non--relaxed} single phonon bath.

\noindent The transport properties of such a system have been
theoretically analyzed both in the sequential tunneling and in the
cotunneling regime~\cite{boese,mitra,koch4,haupt,koch6}.
The behavior strongly differs with respect to the one obtained for a
system coupled to an equilibrated localized phonon mode~\cite{mitra}.
In the sequential tunneling regime, besides the negative differential
conductance mentioned above, the zero--frequency shot noise $S$ has
been found to be even more sensitive to the bath
properties~\cite{mitra}. In the case of strong coupling to an out of
equilibrium phonon bath, a super--Poissonian shot noise $S>2eI$ ($-e$
is the electron charge, $I$ the corresponding current) has been
predicted~\cite{koch4}. On the other hand, for a thermal bath,
sub--Poissonian shot noise $S\leq 2eI$ is always
found~\cite{haupt}. In the cotunneling regime,
vibrational absorption sidebands occur within the Coulomb--blockaded
regime, which disappear for the case of a single bath in thermal
equilibrium~\cite{koch4,koch6,leroy}. This further confirms the importance of
such a unequilibrated phonon bath to describe experimental systems.

In order to consider the coherent dynamics of the quantum dot with the
single mode bath one can derive a generalized master equation within
the Born--Markov approximation for the reduced density matrix of the
system {\em and} the localized bath, by tracing out the leads bath.
Such an approach has been employed already in the past for
describing systems such as a metallic~\cite{armour2} or a
superconducting single electron transistor~\cite{armourmm}.

\noindent Since the transport properties 
 have been already studied in great details, we will
concentrate on the steady state out of equilibrium {\em phonon
distribution} of the bath ($ii$) induced by the tunneling electrons,
which in the past has received much less attention~\cite{mitra,koch5,armourmm,merlo}. 
In particular, we will analyze the phonon Fano factor
\begin{equation}
\label{eq:fano}
F_{ph}=\frac{\mathrm{var}(l)}{\langle l\rangle}=\frac{\langle l^{2}\rangle-\langle l\rangle^{2}}{\langle l\rangle}\,,
\end{equation}
defined as the ratio between the variance of the phonon occupation
number $\mathrm{var}(l)$ and the average occupation number $\langle
l\rangle$. The Fano factor brings information about the statistics of
the single phonon bath mode. For a thermally equilibrated bath, one
always obtains a super--Poissonian Fano factor $F_{ph}>1$. Such a
super--Poissonian value is typical for ``classical'' boson
distributions, as for example for photons in classical
light~\cite{loudon}. As it will be shown later, in most of the
transport regimes, out of equilibrium bath distributions arise which,
although being strongly different from a thermal distribution, display
a super--Poissonian character. However, it is possible to find
particular transport regimes in which $F_{ph}<1$, corresponding to
peculiar bath distributions~\cite{armourmm,merlo}. This case is
analogous to that of the reduced photon fluctuation found in quantum
non--classical radiation~\cite{kimble,itano}.

Our main results are the following. In the regime where the single
bath mode oscillations are much faster than the average electron dwell
time, we adopt the rotating wave approximation and we investigate the
phonon Fano factor. We identify the transport regimes where a
sub--Poissonian single mode bath occurs for the case of a single
occupancy of the dot. For a double occupancy, additional transport
regions in which $F_{ph}<1$ occurs. We evaluate the ``stability
boundary'' of the sub--Poissonian bath as a function of tunnel
barriers asymmetry and coupling strength with the bath mode, finding
that $F_{ph}<1$ is destroyed for increasing damping of the single mode
bath. Finally, we present some preliminary results on the effects of
the coherent phonon bath dynamics occurring when the oscillation time
of the localized mode is of the same order or smaller than the
electron tunneling time.

\section{Model and methods}
\subsection{The system and its environments}
\label{sec:sysenv}
We model the system as a small quantum dot, with an average level
spacing of the same order than the charging energy, 
described as a spin--degenerate single level with on--site Coulomb
repulsion~\cite{schoen}
\begin{equation}
\label{eq:model1}
H_{s}=\xi n-en\frac{C_{g}}{C}V_{g}+\frac{U}{2}n(n-1)\, .
\end{equation} 
Here, $\xi$ is the single energy level,
$n=\sum_{\sigma=\pm}n_{\sigma}$ is the total occupation number with
$n_{\sigma}=d^{\dagger}_{\sigma}d_{\sigma}$ the occupation of spin
$\sigma=\pm 1$ (units $\hbar/2$), and $d_{\sigma}$
($d^{\dagger}_{\sigma}$) are the  fermionic annihilation (creation)
operators. The second term in Eq.~(\ref{eq:model1}) is the
coupling to an external gate voltage $V_{g}$, with $-e$ the electron
charge, $C_{g}$ the gate capacitance, and $C=e^2/U$ the dot
capacitance.  Choosing as a reference level $\xi=U/2$ one has
$H_{s}=\epsilon n+Un(n-1)/2$ with $\epsilon=U(1-2n_{g})/2$ written in
terms of the number of charges induced by the gate
$n_{g}=C_{g}V_{g}/e$. Note that the choice of $\xi$ determines the
values of $n_{g}$ for which resonance between the $n$, $n+1$ states
occur: a different choice of reference for $\xi$ would simply
result in a shift of $n_{g}$.

The system is coupled to two dissipative baths with Hamiltonians ($\hbar=1$)
\begin{equation}
\label{eq:sbm}
H_{b}^{(1)}=\omega_{0}\ b^{\dagger}b\,;\;\qquad 
H_{b}^{(2)}=\!\!\!\!\sum_{\alpha=L,R,k,\sigma=\pm}\varepsilon_{k}\ c^{\dagger}_{\alpha,k,\sigma}c_{\alpha,k,\sigma}\, .
\end{equation}
Here, $H_{b}^{(1)}$ describes the harmonic oscillator single bath mode
with frequency $\omega_{0}$, and $H_{b}^{(2)}$ represents the external
left ($L$) and right ($R$) leads of noninteracting electrons with
fermionic operators $c_{\alpha,k,\sigma}$ and
$c^{\dagger}_{\alpha,k,\sigma}$. The localized phonon mode is
undamped, a discussion of possible damping effects is deferred to
Sec.~\ref{sec:relax}. The leads are assumed in thermal equilibrium
with respect to their electrochemical potential $\mu_{L,R}=\mu_{0}\pm
eV/2$, where $V$ is the applied voltage, and $\mu_{0}$ is the
reference chemical potential. The noninteracting Fermi leads can be
mapped onto a ohmic dissipative bath within the Caldeira Leggett
formalism~\cite{weiss}. More detailed models, involving a
anharmonic~\cite{koch5} or distorted~\cite{nowack} oscillator or the
presence of interacting Luttinger liquid leads~\cite{mitralut}, which
have been studied in the past especially concerning their transport
properties, will not be addressed here.

The coupling of the system with the single-mode bath is linear in the
oscillator coordinate and in the effective charge $n-n_{g}$ on the dot
\begin{equation}
\label{eq:coupsbm}
H_{sb}^{(1)}=\sqrt{\lambda}\omega_{0}(b+b^{\dagger})(n-n_{g})\,.
\end{equation}
This describes realistic situations such as a single electron transistor
capacitively coupled to a vibrating, charged
gate~\cite{knobel,armour1}.
\noindent The tunneling Hamiltonian
\begin{equation}
\label{eq:tunnel}
H_{sb}^{(2)}=\sum_{\alpha=L,R}t_{\alpha}\sum_{k,\sigma=\pm}c^{\dagger}_{\alpha,k,\sigma}d_{\sigma}+h.c.\, .
\end{equation}
represents the coupling of the dot with the external leads, through
tunneling amplitudes $t_{L}$ and $t_{R}$. In the following, a trace
over the leads degrees of freedom will be performed.

The full dynamics of the single mode will be retained, and
its degrees of freedom will {\em not} be traced away. It is possible
to diagonalize exactly the Hamiltonian representing the dot coupled to 
single phonon mode
\begin{equation}
\label{eq:hlambda}
H_{\lambda}=H_{s}+H_{b}^{(1)}+H_{sb}^{(1)}\,,
\end{equation}
by means of the Lang--Firsov polaron transformation~\cite{mitra}, with
generator $\mathcal U=\exp{[-\eta(b-b^{\dagger})]}$, and
$\eta=\sqrt{\lambda}(n-n_{g})$. The transformed operators, denoted by
an overbar
$\bar{\mathcal{O}}=\mathcal{U}\mathcal{O}\mathcal{U}^{\dagger}$, are
$\bar{b}=b-\eta$ and
$\bar{d}_{\sigma}=d_{\sigma}\exp{\left[\sqrt{\lambda}(b-b^{\dagger})\right]}$,
while $\bar{n}=n$. Upon transformation, we have
\begin{equation}
\label{eq:model3}
\bar{H}_{\lambda}=\bar{\epsilon}n+\frac{\bar{U}}{2}n(n-1)+\omega_{0}l\, ,
\end{equation}
with renormalized level position
$\bar{\epsilon}$ and Coulomb repulsion
$\bar{U}$
\begin{equation}
\bar{\epsilon}=\frac{\bar{U}}{2}(1-2n_{g});\qquad \bar{U}=U-2\lambda\omega_0\,.
\label{ubar}
\end{equation}
The eigenvectors  of Eq.~(\ref{eq:model3}) are denoted by $|n,l\rangle$,
with energy $E_{n,l}$. Note that the polaronic renormalization of the
Coulomb interaction may lead to two qualitatively different physical
scenarios for $\bar{U}\lessgtr0$ or equivalently $\lambda\gtrless
U/2\omega_{0}$.  The regime $\bar{U}<0$, where the {\em single
  occupation} is always forbidden and the sequential transport is
blocked, has been considered by several
authors~\cite{koch2,cornaglia1,mravlje}.  In the
present paper, we will consider the opposite case with $\bar{U}>0$
treating the sequential tunneling regime.

\noindent Finally, the transformed tunneling Hamiltonian is 
\begin{equation}
\label{eq:tunnel2}
\bar{H}_{sb}^{(2)}=\sum_{\alpha=L,R}t_{\alpha}e^{\sqrt{\lambda}(b-b^{\dagger})}\sum_{k,\sigma=\pm}c^{\dagger}_{\alpha,k,\sigma}d_{\sigma}+h.c.\,,
\end{equation}
with an the explicit dependence on the oscillator variables.
\subsection{Generalized master equation}
\label{sec:master}

The dynamics of the dot and single bath mode can be described in terms
of the reduced density matrix ${\rho}(t)$, defined as the trace over
the {\em leads} bath of the total density matrix $\rho_{tot}(t)$:
${\rho(t)}=\mathrm{Tr}_{l}\{{\rho}_{tot}(t)\}$. In the following we
will mainly work {\em in the polaron frame} with the density matrix
denoted by $\bar{\rho}(t)$, the evolution in the original frame being
easily traced back via the canonical transformation
$\rho(t)=\mathcal{U}^{\dagger}(t)\bar{\rho}(t)\mathcal{U}(t)$.

We will consider the regime of weak tunneling, treating
$\bar{H}_{sb}^{(2)}$ in Eq.~(\ref{eq:tunnel}) to lowest order, and
assume the characteristic memory time of the leads much shorter than
the response time of the dot interacting with the localized phonon
bath.  This allows to treat the dynamics in the Born-Markov
approximation~\cite{cohen} leading to a generalized master equation
(GME). Assuming a factorized total density
$\bar{\rho}_{tot}(t)=\bar{\rho}(t)\otimes\rho_{l}(0)$ with the leads
in thermal equilibrium with respect to corresponding chemical
potential $\mu_{L,R}$: $\rho_{l}(0)=\rho_{L}(0)\otimes\rho_{R}(0)$,
the time evolution in the interaction picture (denoted with the
subscript ``$I$'') is
\begin{eqnarray}
\label{eq:master3}
&\dot{\bar{\rho}}_{I}(t)&=-\sum_{\sigma=\pm}\int_{0}^{\infty}{\rm
  d}\tau\left\{[Q_{I,\sigma}(t),Q^{\dagger}_{I,\sigma}(t-\tau)\bar{\rho}_{I}(t)]K^{+}(\tau)\right.\nonumber\\
&-&[Q_{I,\sigma}(t),\bar{\rho}_{I}(t)Q^{\dagger}_{I,\sigma}(t-\tau)]K^{-}(\tau)
+[Q^{\dagger}_{I,\sigma}(t),Q_{I,\sigma}(t-\tau)\bar{\rho}_{I}(t)]K^{-}(-\tau)\nonumber\\
&-&\left.[Q^{\dagger}_{I,\sigma}(t),\bar{\rho}_{I}(t)Q_{I,\sigma}(t-\tau)]K^{+}(-\tau)\right\}\, .
\end{eqnarray}
Here, we defined the operator
$Q_{\sigma}=e^{\sqrt{\lambda}(b-b^{\dagger})}d_{\sigma}$ and the leads
correlation functions
\begin{equation}
\label{eq:correlations}
K^{\pm}(\tau)=\sum_{\alpha,k}|t_{\alpha}|^{2}e^{i\varepsilon_{k}\tau}f^{\pm}_{\alpha}(\varepsilon_{k})\, ,
\end{equation}
with
$f^{+}_{\alpha}(\varepsilon)=1/\{1+\exp{[\beta(\varepsilon-\mu_{\alpha})]}\}$
the Fermi distribution of lead $\alpha$ and
$f^{-}_{\alpha}(\varepsilon)=1-f^{+}_{\alpha}(\varepsilon)$. 
Note that the correlation
function $K^{+}(\tau)$ can be cast into the form~\cite{sassetti,cavaliere}
\begin{equation}
\label{eq:dissi1}
K^{+}(\tau)=\nu\sum_{\alpha}|t_{\alpha}|^{2}e^{i\mu_{\alpha}\tau}\omega_{c}e^{-W(\tau)}\, ,
\end{equation}
where $\nu$ is the leads density of states and 
\begin{equation}
\label{eq:dissi2}
W(\tau)=\int_{0}^{\infty}\mathrm{d}\omega\ \frac{J(\omega)}{\omega^2}\left\{[1-\cos{(\omega\tau)}]
\coth{\left(\frac{\beta\omega}{2}\right)}+i\sin(\omega\tau)\right\}\,,
\end{equation}
represents the dissipative kernel of a set of harmonic oscillators
with ohmic spectral density $J(\omega)=\omega e^{-\omega/\omega_{c}}$,
and frequency cutoff $\omega_c$~\cite{weiss}.
Expression~(\ref{eq:dissi1}) clarifies the connection between the
electronic leads and the bosonic modes of the Caldeira Leggett model. 

The above approximations allow to describe the dynamics of the dot in
the so called sequential tunneling regime.  They are typically valid
for temperatures larger than the level broadening induced by tunneling
$k_B T\gg
\Gamma_{L}+\Gamma_{R}$, with
$\Gamma_{\alpha}=2\pi\nu|t_{\alpha}|^{2}$. At lower temperatures,
coherences between the leads and the dot play a crucial
role~\cite{mitra}. These effects will be not discussed in this work.

It is convenient  to project Eq.~(\ref{eq:master3}) on the eingenstates of
the Hamiltonian~(\ref{eq:model3}) and perform the time integrations
with the aid of $\int_{0}^{\infty}\mathrm{d}\tau\
e^{iE\tau}=\pi\delta(E)+i\mathrm{P.V.}(1/E)$. In the following, terms
stemming from the principal value are neglected since they lead to
small corrections in the perturbative
regime~\cite{armour2,schoen2}. Denoting with
$\bar{\rho}_{I,qq'}^{n}(t)=\langle n,q|\bar{\rho}_{I}(t)|n,q'\rangle$
the GME has the compact form (here and in the following, $0\leq
n\leq 2$ unless stated otherwise)
\begin{eqnarray}
\label{eq:gme_inter}
\!\!\!\!\!\!\!\!\!\!\!\!\!\!\!\!\!\dot{\bar{\rho}}_{I,qq'}^{n}(t)&=&\sum_{k=\pm 1\atop 0\leq n+k\leq 2}\sum_{p,p'}\left\{z_{n+k}X^{n+k,n}_{q'p'}X^{n+k,n}_{qp}(\gamma^{n+k,n}_{pq}
+\gamma^{n+k,n}_{p'q'})\bar{\rho}^{n+k}_{I,pp'}(t)e^{-i(\omega_{q'q}+\omega_{pp'})t}\right.\nonumber\\
&-&\left. z_{n}X^{n,n+k}_{p',p}\gamma^{n,n+k}_{pp'}[X^{n,n+k}_{p'q}\bar{\rho}^{n}_{I,pq'}(t)
e^{-i\omega_{pq}t}+X^{n,n+k}_{p'q'}\bar{\rho}^{n}_{I,qp}(t)e^{-i\omega_{q'p}t}]\right\}\, .
\end{eqnarray}
Here we defined $\omega_{pp'}=\omega_{0}\cdot (p-p')$, with the
indexes $p,p',q,q'$ that run over $[0,\infty)$. The factors
$z_{0}=z_{2}=2$, $z_{1}=1$ arise due to the spin degeneracy of the
state $n=1$. Furthermore, it is
\begin{equation}
\label{eq:gammas}
\hspace{-1.5cm}\gamma_{pq}^{n,n+1}=\sum_{\alpha}\frac{\Gamma_{\alpha}}{2}f_{\alpha}^{+}(E_{n+1,q}-E_{n,p})\quad\quad\quad\gamma_{pq}^{n+1,n}=\sum_{\alpha}\frac{\Gamma_{\alpha}}{2}f_{\alpha}^{-}(E_{n+1,p}-E_{n,q})
\end{equation}
and we have introduced
\begin{equation}
X^{n+1,n}_{qq'}=X_{qq'}\quad\quad\quad X^{n,n+1}_{qq'}=X_{q',q}
\end{equation}
with $X_{qq'}=\langle
n,q|e^{\sqrt{\lambda}(b-b^{\dagger})}|n,q'\rangle$ the Franck--Condon
factors, given by
\begin{equation}
\label{eq:franck}
X_{qq'}=e^{-\lambda/2}\sqrt{\frac{q_{<}!}{q_{>}!}}L_{q_{<}}^{|q'-q|}(\lambda)\left[\mathrm{sgn}(q'-q)\sqrt\lambda\right]^{|q'-q|}\, ,
\end{equation}
where $q_{<}=\mathrm{min}\{q,q'\}$, $q_{>}=\mathrm{max}\{q,q'\}$ and
$L_{\mu}^{\nu}(\lambda)$ is the generalized Laguerre polynomial. They satisfy the relation
$X_{qq'}^{2}=X_{q'q}^{2}$. 

\noindent In the Schr\"odinger representation,
Eq.~(\ref{eq:gme_inter}) becomes 
\begin{eqnarray}
\label{eq:gme_schro}
\!\!\!\!\!\!\!\!\!\!\!\!\!\!\dot{\bar{\rho}}_{qq'}^{n}(t)&=&-i\omega_{qq'}\bar{\rho}_{qq'}^{n}(t)+\!\!\!\sum_{k=\pm 1\atop 0\leq n+k\leq 2}\!\!\sum_{p,p'}
\left\{z_{n+k}X^{n+k,n}_{q'p'}X^{n+k,n}_{qp}(\gamma^{n+k,n}_{pq}+\gamma^{n+k,n}_{p'q'})\bar{\rho}^{n+k}_{pp'}(t)\right.\nonumber\\
&-&\left. z_{n}X^{n,n+k}_{p'p}\gamma^{n,n+k}_{pp'}[X^{n,n+k}_{p'q}\bar{\rho}^{n}_{pq'}(t)+X^{n,n+k}_{p'q'}\bar{\rho}^{n}_{qp}(t)]\right\}\, .
\end{eqnarray}
The GME~(\ref{eq:gme_schro}) is an infinite set of coupled linear
differential equations which cannot be solved analytically under
general conditions.  We numerically solve the system truncating the
harmonic oscillator Hilbert space to increasingly larger sizes
$q,q'\lesssim 70$, until convergence is achieved.

\subsection{Rotating wave approximation}
\label{sec:rwa}
In the regime of fast vibrational motion of the localized bath,  
$\omega_{0}\gg\Gamma_{L},\Gamma_{R}$, one can perform
the rotating wave approximation (RWA), neglecting terms in
Eq.~(\ref{eq:gme_inter}) with an explicit oscillatory exponential time
dependence~\cite{cohen}. In the Schr\"odinger
representation, this leads to a GME of the form
\begin{eqnarray}
\label{eq:gme_rwa}
\!\!\!\!\!\!\!\!\!\!\!\!\!\!\!\!\!\!\!\!\!\!\!\dot{\bar{\rho}}_{qq'}^{n}(t)&=&-i\omega_{qq'}\bar{\rho}_{qq'}^{n}+
\sum_{k=\pm 1\atop 0\leq n+k\leq 2}\left\{
\sum_{p=-q_{<}}^{\infty}z_{n+k}X^{n+k,n}_{q,q+p}X^{n+k,n}_{q',q'+p} \bar{\rho}^{n+k}_{q+p,q'+p}\right.(\gamma^{n+k,n}_{q+p,q}+\gamma^{n+k,n}_{q'+p,q'})\nonumber\\
\!\!\!\!\!\!\!\!\!&&\left.-z_{n}\bar{\rho}^{n}_{q,q'}\sum_{p=0}^{\infty}
\left[(X^{n,n+k}_{pq'})^{2}\gamma^{n,n+k}_{q'p}+(X^{n,n+k}_{pq})^{2}\gamma^{n,n+k}_{qp}\right]\right\}\, .
\end{eqnarray}
Note that in the RWA the density matrix elements
$\bar{\rho}_{qq'}^{n}$ and $\bar{\rho}_{ll'}^{n'}$ are coupled only if
$l'-l=q'-q$. This implies that the diagonal elements
$\bar{\rho}_{qq}^{n}(t)$ are decoupled from the off-diagonal ones.  In
addition, the non diagonal elements vanish in the stationary regime,
$\bar{\rho}^{n}_{l,l'\neq l}(t\to\infty)\to 0$, as a difference with
the coherent regime (see section 4.2).  Hence, within the RWA all the
stationary properties are determined by the diagonal occupation
probabilities $\bar{\rho}_{qq}^{n}\equiv \bar{P}_{nq}$ and
Eq.~(\ref{eq:gme_rwa}), in the stationary limit, assumes the form of a
standard rate equation
\begin{eqnarray}
\label{eq:master_rwa}
&&z_{n}\bar{P}_{nq}\sum_{k=\pm 1\atop 0\leq n+k\leq 2}\sum_{p=0}^{\infty}
\Gamma_{qp}^{n,n+k}-\sum_{k=\pm 1\atop 0\leq n+k\leq 2}\sum_{p=0}^{\infty}z_{n+k}\bar{P}_{n+k,p}\Gamma^{n+k,n}_{pq}=0\\
\label{rates}
&&\Gamma^{nn'}_{pq}=\Gamma^{nn'}_{L,pq}+\Gamma^{nn'}_{R,pq}=2X_{p,q}^{2}\gamma^{nn'}_{pq}
\end{eqnarray}
with $\Gamma^{nn'}_{pq}$ the tunneling rates for the transition
$|n,p\rangle\to|n',q\rangle$ and $\Gamma_{\alpha,pq}^{nn'}$ the
tunneling rate on barrier $\alpha$. We specify here the energy regions
where tunneling processes with rates $\Gamma^{nn'}_{pq}$ are allowed
(i.e.  $\Gamma^{nn'}_{pq}\neq 0$) for temperatures low enough that the
Fermi function can be approximated as a step, $T\ll\omega_{0}/k_{B}$.
From the definition~(\ref{eq:gammas}) follows
\begin{eqnarray}
\label{eq:open1}
\Gamma_{pq}^{n,n+1}\neq 0\ &\Rightarrow&\ E_{n+1,q}-E_{n,p}\pm eV/2\leq 0\\
\label{eq:open2}
\Gamma_{pq}^{n+1,n}\neq 0\ &\Rightarrow&\ E_{n+1,p}-E_{n,q}\pm eV/2\geq 0\, ,
\end{eqnarray}
with the $+$ ($-$) sign for a tunneling event on the right (left)
barrier. The Franck Condon factors in the tunneling
rates Eq.(\ref{rates}) are responsible for two important  effects:

\noindent ($i$) suppression of all the tunneling rates by a factor
$e^{-\lambda}$,

\noindent ($ii$)  non--trivial dependence on the phonon
indexes $p$, $q$. 

\noindent In particular, for moderate interactions $\lambda\lesssim
1$, transitions which conserve or change slightly the phonon number
have the largest rate and, among these, the ones involving small $p$
and $q$ are dominant. Vice versa, for $\lambda\gg 1$, transitions with
a large change in the phonon number are favored.  These
considerations will play an important role in explaining both the
transport properties and the characteristics of the single mode bath.

Finally, the stationary current can be expressed in terms of the
occupation probabilities as follows
\begin{equation}
I=-e\sum_{n=0,1}\sum_{q,q'}\left[z_n\bar{P}_{nl}\Gamma_{L,qq'}^{n,n+1}-z_{n+1}\bar{P}_{n+1,l}\Gamma_{L,qq'}^{n+1,n}\right]\, ,
\label{eq:current}
\end{equation}
note that $I$ is independent on the position.

\noindent In order to obtain the stationary solution of the phononic populations,
we numerically solved Eq.~(\ref{eq:master_rwa}) for a sufficiently
large set of harmonic oscillator states with $q\lesssim 60$, which
guarantees a fair convergence. A discussion of the range of validity of
this approximation will be presented in Section~\ref{sec:coherence}.
\section{Results} 
In the following we will present several effects induced by the
coupling of the quantum dot with the unequilibrated single bath mode.
We will consider the regime $k_{B}T\ll\omega_{0}$ where the quantization effects of the 
phonon mode are more visible.
\begin{figure}[h]
\begin{center}
\includegraphics[width=13cm,keepaspectratio]{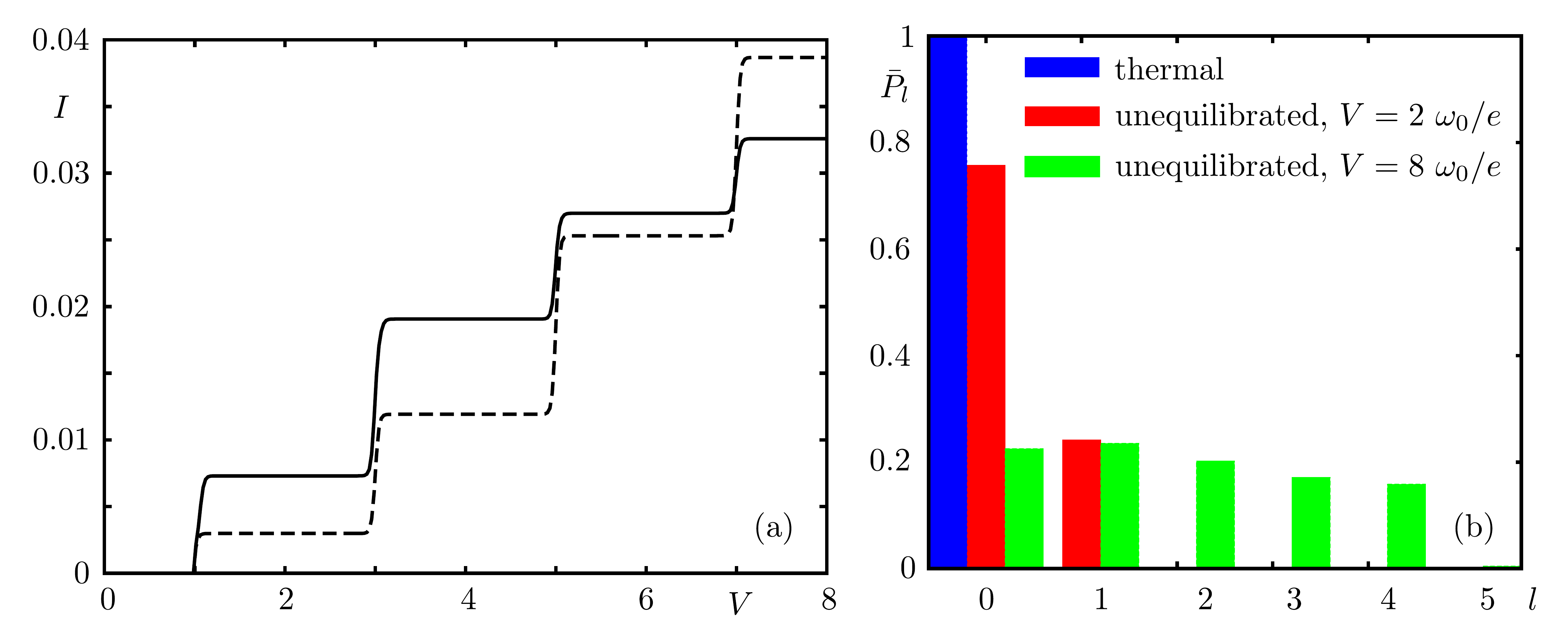}
\caption{(a) Current $I$ (units $-e\Gamma_{L}$) as a function of $V$
  (units $\omega_{0}/e$) for an unequilibrated (solid) or a thermal
  (dashed) phonon bath, for $n_{g}=0.525$, $A=0.01$, $\lambda=3$,
  $k_{B}T=0.01\ \omega_{0}$, $\bar{U}=\ 20\ \omega_{0}$ and
  $\Gamma_{L}=10^{-3}\omega_{0}$. (b) Phonon distributions
  $\bar{P}_{l}$, as a function of $l$, for the cases of a thermal bath (blue) and an
  unequilibrated bath calculated for $eV=2\ \omega_{0}$ (red) and
  $eV=8\ \omega_{0}$ (green), all other parameters as in panel (a).}
\label{fg:fig0}
\end{center}
\end{figure}
Figure~\ref{fg:fig0}(a) shows the current $I$, numerically evaluated
from Eq.(\ref{eq:current}) in the RWA, as a function of voltage in the presence of an asymmetry $A=|t_{R}|^{2}/|t_{L}|^{2}<1$. Two
different conditions of the single bath mode are considered: a completely
out of equilibrium distribution $\bar{P}_{l}$, determined by solving
the rate equations~(\ref{eq:master_rwa}), and a thermal distribution
$\bar{P}_{l}^{(th)}$ 
\begin{equation}
\label{eq:thermal}
\bar{P}_{l}=\sum_{n}\bar{P}_{nl}\,;\qquad 
\bar{P}_{l}^{(th)}=e^{-l\beta\omega_{0}}(1-e^{-\beta\omega_0})\, ,
\end{equation}
and $\beta= (k_{B}T)^{-1}$.  For $V<\omega_{0}/e$, the dot is in the
Coulomb blockade regime. For increasing $V$, steps in the current
reflect phonon excitations in the bath. The solid line is the result
with $\bar{P}_{l}$, the dashed line displays the current obtained
imposing a thermal distribution of the bath mode. The corresponding
phonon distributions $\bar{P}_{l}$ deviates considerably from
$\bar{P}_{l}^{(th)}$, as shown in Fig.~\ref{fg:fig0}(b).  While
$\bar{P}_{l}^{(th)}$ at $T\ll\omega_0/k_{B}$ is almost concentrated at
$l=0$, the unequilibrated distributions are broader.

The behavior of the current is qualitatively different, depending on
the bath conditions. In particular, for $V<5\ \omega_{0}/e$ the
current in the equilibrated case is smaller than for an unequilibrated
phonon bath while for $V>\ 7\omega_{0}/e$ the situation is
reversed~\cite{mitra,haupt}. This fact can be explained as follows.
For $\lambda>1$, as in the case of Fig.~\ref{fg:fig0}(a), the dominant
transition involves $l=0\to\mathrm{int}(\lambda)$~\cite{haupt}, as a
consequence of the Franck Condon factors~(\ref{eq:franck}). For the
parameters in the figure, it corresponds to the transition $l=0\to 3$,
which is open only for $V>5\omega_{0}$ as can be checked by inspecting
Eqs.~(\ref{eq:open1}),(\ref{eq:open2}). A large occupation of the
states with $l=0$ leads therefore to an increase of the current. This
explains why, for large voltages $V>7\omega_{0}/e$, in the case of a
thermal distribution $\bar{P}_{l}^{(th)}$ one obtains a larger current
with respect to the unequilibrated case where a {\em broader} phonon
distribution implies a {\em smaller} $\bar{P}_{0}$~\cite{haupt}. Vice
versa, for $V<5\ \omega_{0}/e$ only small tunneling rates with
$\Delta l\leq 2$ are open, therefore the larger the number of
transport channels, the larger is the current. In the case of a
thermal bath only transitions originating from $l=0$ can contribute to
the current, while in the unequilibrated case also transitions
starting from $l=1$ and $l=2$ states contribute, resulting in a higher
current with respect to the equilibrated case.

Also the current fluctuations  display
deviations (not shown here): in the unequilibrated case a super--Poissonian {\em
current noise} appears, while for a thermal bath one obtains a
sub--Poissonian noise~\cite{koch4,haupt}. These topics have been
studied in great details~\cite{boese,mitra,koch4,haupt,koch6}, thus we will
not repeat here the analysis of transport properties induced by
the unconventional localized mode.

\noindent Instead, we will focus on {\em the properties of the single
  bath mode}, and we will characterize the out of equilibrium phonon
distribution induced by the current flow. In particular, we will
analyze the Fano factor. The averages involved in Eq.~(\ref{eq:fano})
are equivalently defined in the original frame or in the polaron one
as
$\langle\mathcal{O}\rangle=\mathrm{Tr}\{\mathcal{O}\rho\}=\mathrm{Tr}\{\bar{\mathcal{O}}\bar{\rho}\}$
with the trace performed over the degrees of freedom of the system and
of the single bath mode. Since operators in the polaron frame are
expressed in terms of the ones in the original frame, it will be
particularly useful to introduce the hybrid average
$\langle\mathcal{O}\rangle_{\bar{\rho}}=\mathrm{Tr}\{\mathcal{O}\bar{\rho}\}$.
With respect to this average the above occupation number and variance
are given by ($\eta=\sqrt{\lambda}(n-n_{g})$)
\begin{eqnarray}
\label{eq:avgl}
\langle l\rangle&=&\langle l\rangle_{\bar{\rho}}-\langle \eta x\rangle_{\bar{\rho}}+\langle \eta^{2}\rangle_{\bar{\rho}}\\
\label{eq:avgl2}
\langle l^{2}\rangle&=&\langle l^{2}\rangle_{\bar{\rho}}+\langle \eta^{4}\rangle_{\bar{\rho}}+\langle \eta^{2}x^{2}\rangle_{\bar{\rho}}+2\langle \eta^{2}l\rangle_{\bar{\rho}}-2\langle \eta^{3}x\rangle_{\bar{\rho}}-\langle \eta\{l,x\}\rangle_{\bar{\rho}}\, ,
\end{eqnarray}
where $\{a,b\}=ab+ba$ and $x=b+b^{\dagger}$. Note that the Fano factor
can also be connected to the Mandel parameter
$\mathcal{Q}=F_{ph}-1$~\cite{mandel}, introduced in quantum optics to
discriminate the sub--Poissonian statistics and photon
antibunching~\cite{kimble,itano}.

We point out here that in the case of a thermally equilibrated bath,
with a diagonal density matrix Eq.~(\ref{eq:thermal}) right side, one has
$F_{ph}>1$~\footnote{Note that in the thermal regime, also the density
matrix {\em in the original frame} is diagonal if expressed on the
basis of the eigenstates of $H_{\lambda}$, given by
$U^{\dagger}|n,l\rangle$.}. As we will see below, deviations from an
equilibrated bath usually result in  super--Poissonian
distributions. However, in special transport regimes it is possible to
obtain peculiar distributions with a sub--Poissonian character.
\subsection{Single occupancy}
\label{sec:nodouble}

We start from the case $\bar{U}\gg\omega_0$ within the RWA.
In this regime the dot, around the resonance condition $n_g\approx
1/2$, has a single occupancy with $n=0,1$ and shows qualitatively
similar behavior to that discussed in Ref.~\cite{merlo}, where
$\bar{U}\to\infty$. Within the RWA, the Fano factor is simplified since
some averages in~(\ref{eq:avgl}) and (\ref{eq:avgl2}) vanish due to the
diagonal form of the density matrix.  It is convenient to express
$F_{ph}$ in terms of the numerator of the Mandel parameter
$F_{ph}=1+\mathcal{Q}_{num}/\langle l\rangle$ where
\begin{eqnarray}
\label{eq:fullQ}
\mathcal{Q}_{num}&=&\mathcal{Q}_{\bar{\rho}}+4\langle
\eta^{2}l\rangle_{\bar{\rho}}-2\langle
\eta^{2}\rangle_{\bar{\rho}}\langle l\rangle_{\bar{\rho}}+\langle
\eta^{4}\rangle_{\bar{\rho}}-\langle
\eta^{2}\rangle_{\bar{\rho}}^{2}\\
\label{eq:intrQ} 
\mathcal{Q}_{\bar{\rho}}&=&\langle
l^{2}\rangle_{\bar{\rho}}-\langle l\rangle_{\bar{\rho}}^{2}-\langle
l\rangle_{\bar{\rho}}\,.
\end{eqnarray}
Here, we have separated a contribution, $\mathcal{Q}_{\bar{\rho}}$, depending only on
the phonon distribution, and terms related to charge fluctuations on the
dot (terms in $\eta^2$ and $\eta^4$).  In general, one has $\mathcal{Q}_{num}\gtrless 0$ and
$\mathcal{Q}_{num}>\mathcal{Q}_{\bar{\rho}}$~\cite{merlo}. 
The necessary condition to have $F_{ph}<1$ is $\mathcal{Q}_{\bar{\rho}}<0$. 
However, this does not guarantee sub-Poissonian behavior since the variance
$\langle\eta^{4}\rangle_{\bar{\rho}}-\langle
\eta^{2}\rangle_{\bar{\rho}}^{2}$, being definite positive, could drive
$F_{ph}>1$.

\noindent The quantity
$\mathcal{Q}_{\bar{\rho}}$ in (\ref{eq:intrQ}) 
is expressed in terms of the phonon distribution $\bar{P}_{l}$ in Eq.~(\ref{eq:thermal})
as follows
\begin{equation}
\mathcal{Q}_{\bar{\rho}}=\sum_{l}\bar{P}_{l}l(l-1) -\left(\sum_{l}\bar{P}_{l}l\right)^2\,.
\end{equation}
A sub-Poissonian distribution can originate from peculiar phonon populations characterized by
$\mathcal{Q}_{\bar{\rho}}<0$. This may be achieved under the following
{\em selective}  condition
\begin{equation}
\bar{P}_{0}\approx\bar{P}_{1}\gg\bar{P}_{l\geq 2}\,.
\label{eq:selective}
\end{equation}
Phonon distributions satisfying the above condition in the following will be referred to as 
"selective populations". They will play an important role in
inducing a ``non--classical'' phonon gas.  We now analyze the parameter
regime where this condition can be achieved.  
\begin{figure}[h]
\begin{center}
\includegraphics[width=15cm,keepaspectratio]{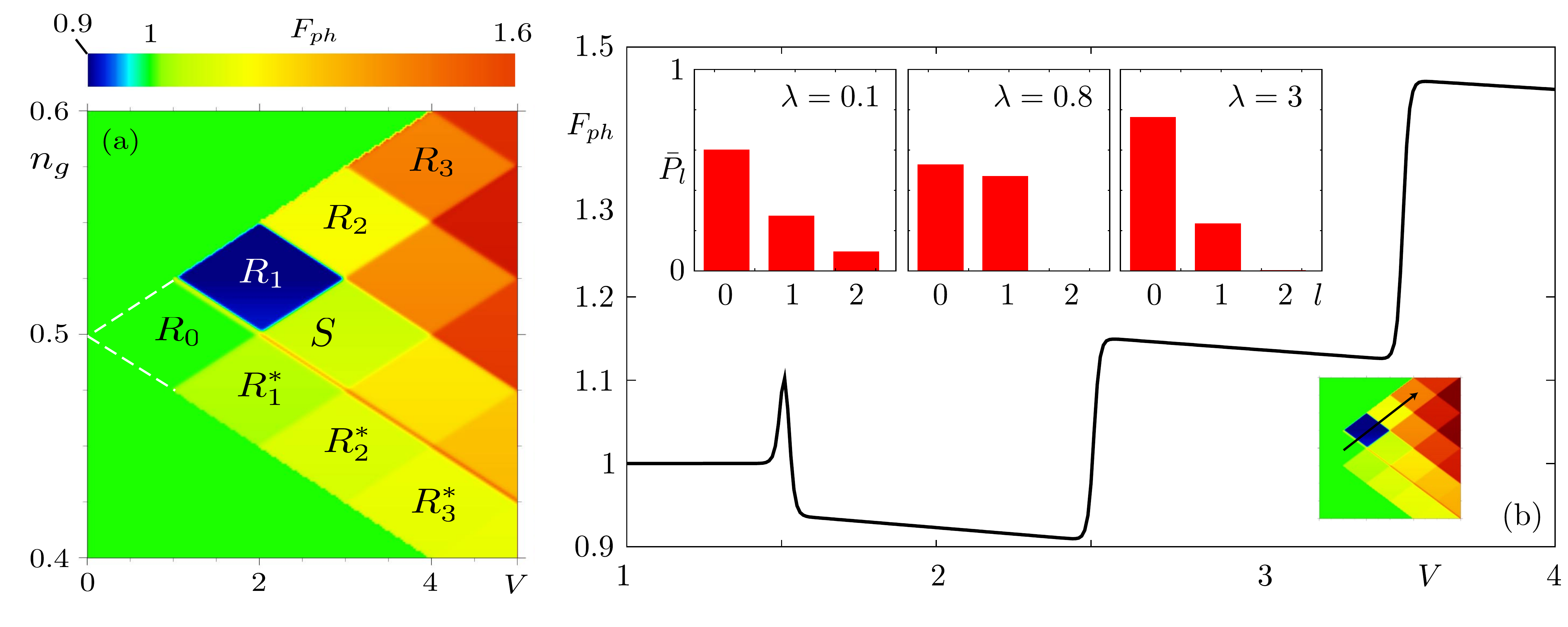}
\caption{(a) Color map of the numerically calculated phonon Fano
  factor $F_{ph}$ as a function of $V$ (units $\omega_0/e$) and
  $n_{g}$ for $A=0.1$, $\lambda=0.8$, $k_{B}T=0.01\ \omega_{0}$,
  $\bar{U}=\ 20\ \omega_{0}$ and $\Gamma_{L}=10^{-3}\omega_{0}$. (b)
  Main: $F_{ph}$ as a function of $V$ (units $\omega_{0}/e$)
  calculated along the line
  $\bar{U}(1-2n_{g})+eV-\omega_{0}=0$ in the $V$, $n_{g}$
  plane (corresponding to the black arrow in the bottom right
  scheme). Inset: phonon distributions $\bar{P}_{l}$ as a function of
  $l$ for different values of $\lambda$, calculated in the middle of
  region $R_{1}$ ($n_{g}=0.525$, $eV=3\omega_0$); other parameters as
  in (a). For $\lambda=0.1,0.8,3$ (left to right panels) it is
  $F_{ph}\approx 1.2,0.9,1.3$ respectively.}
\label{fg:fig1}
\end{center}
\end{figure}

\noindent Fig.~\ref{fg:fig1}(a) shows a color map of the numerically
calculated $F_{ph}$ as a function of $V$ and $n_{g}$ for ${\bar
U}=20\,\omega_0$. It displays a checkerboard pattern where each
region is characterized by the activation of transport channels
involving transitions between different phonon states. The regions $R_{p}$ and $R_{p}^{*}$ will be particularly important for the following discussions. Their center is at
\begin{equation}
\label{eq:centerRp}
V^{(c)}=(p+1)\frac{\omega_{0}}{e}\quad\quad;\quad\quad n_{g}^{(c)}=\frac{1}{2}\pm\frac{\omega_{0}}{2\bar{U}}p
\end{equation}
with the $+$ ($-$) sign for region $R_{p}$ ($R_{p}^{*}$). As stems
from~(\ref{eq:open1}),(\ref{eq:open2}) the following conditions are
necessary for transitions to be open
\begin{eqnarray}
\label{eq:trans1}
|0,l+q\rangle&\to&|1,l'+q\rangle\ \Leftrightarrow\ 2\omega_{0}(l'-l)+\bar{U}(1-2n_{g})\pm eV\leq 0\\
\label{eq:trans2}
|1,l+q\rangle&\to&|0,l'+q\rangle\ \Leftrightarrow\ 2\omega_{0}(l'-l)-\bar{U}(1-2n_{g})\pm eV\leq 0\, .
\end{eqnarray}
For the values of $V$ and $n_{g}$ within region $R_{p}$, it follows
that the transitions~(\ref{eq:trans1}) are open if $l'\leq l+p$, while
for the transitions~(\ref{eq:trans2}) one needs $l'\leq l$. Similarly,
in regions $R_{p}^{*}$  one finds that for the
transitions~(\ref{eq:trans1}) one has $l'\leq l$ while
for~(\ref{eq:trans2}) it is $l'\leq l+p$.

\noindent Thus, in region $R_{0}$ the transition rates towards excited
phonon states are all closed, therefore only the $l=0$ phonon
state is populated. Here, the solution of the rate equations is
simply given by $\bar{P}_{00}=A/(A+2)$,
$\bar{P}_{10}=1-\bar{P}_{00}$ (for $T\ll\ \omega_{0}/k_{B}$) and one
gets a super-Poissonian behavior
\begin{equation*}
F_{ph}=1+\lambda\frac{2A}{2+A}\frac{(2n_{g}-1)^{2}}{(2+A)n_{g}^{2}-4n_{g}+2}>1\, .
\end{equation*}

\noindent The situation is more interesting in the other regions displayed in
the figure. Indeed, in most of them the out of equilibrium phonon bath
displays a super--Poissonian character.  However, one can note that
sub--Poissonian values, $F_{ph}<1$, are present at low voltages in
region $R_{1}$ as shown also in the main panel of
Fig.~\ref{fg:fig1}(b). Concentrating on $R_{1}$, the selective
population arises if the following conditions are met:

\noindent ($i$) the allowed phonon transitions
$|n,l\rangle\to|n',l'\rangle$ must satisfy $l'\leq 1+l$, which implies
no {\em direct} population of the states with $l\geq 2$ from an
initial state $|n,0\rangle$;

\noindent ($ii$) the asymmetry has to be $A<1$, which suppresses the rate for
tunnel--out transition $|1,l\rangle\to|0,l'\rangle$ since for $V>0$
electrons flow from the left to the right lead.

\noindent Condition ($i$) is precisely realized in region $R_{1}$. By virtue of
this, in order to populate excited phonon states with $l\geq 2$
starting from the phonon ground state $l=0$ within $R_{1}$, one needs
to perform {\em at least} the following sequence of transitions
\begin{equation}
\label{eq:chain}
|0,0\rangle\to|1,1\rangle\to|0,1\rangle\to|1,2\rangle\to\ldots\, .
\end{equation}
This implies that at least $l-1$ tunnel--out events are performed in
$R_{1}$ in order to populate an excited state with $l\geq 1$ and this leads
to $\bar{P}_{l\geq 1}\propto A^{l-1}$.

\noindent Condition ($ii$) and the above result induce a small occupation of the $l\geq 2$ states for $A<1$~\cite{merlo}. The
inset of Fig.~\ref{fg:fig1}(b) shows the phonon distribution in region
$R_1$ for $A=0.1$ and three different coupling strengths. One can
observe that all distributions are markedly out of equilibrium and
very different from a thermal population, see the blue bar in
Fig.~\ref{fg:fig0}(b). In addition, only the case of
intermediate coupling $\lambda=0.8$ (second panel) displays a
selective population as defined above. For small $\lambda$ (first
panel) broad phonon distributions are obtained. On the other hand, for
$\lambda>1$, the phonon distribution gets very narrow with
$\bar{P}_{0}\gg\bar{P}_{l\geq 1}$ and no selective phonon population
as in (\ref{eq:selective}) may be achieved.

\noindent As a result of the competition between asymmetry, coupling
strength and charge fluctuations, it is possible to find $F_{ph}<1$
only in the parameter space characterized by {\em both}
$\lambda\lesssim 1$ {\em and} $A<A_{c}(\lambda)<1$ where
$A_{c}(\lambda)$ is a threshold asymmetry of order unity for
$\lambda\approx 1$~\cite{merlo}. Furthermore, it is possible to
show~\cite{merlo} that similar arguments apply in the region $S$ with
a much stronger asymmetry, while region $R_{1}^{*}$ is similar to
$R_{1}$, inverting the asymmetry: $A\to 1/A$.  

In all other regions $R_{p\geq 2}$ and $R_{p\geq 2}^{*}$, the
transitions $|n,l\rangle\to|n',l'\rangle$ with $l'\leq l+p$ and $p\geq
2$ are active. Now, the transitions $|n,0\rangle\to|n',l'\rangle$ with
$l'\leq p$ {\em directly} populate phonon states with $l'\geq 2$ from
the phonon ground state $|n,0\rangle$. As a result, the mechanism to
obtain a selective population discussed above is no longer effective,
and {\em broad} phonon distributions are attained for $\lambda\lesssim
1$. They neither display a selective population, nor
$F_{ph}<1$. Finally, for $\lambda>1$, narrow distributions arise and
again no $F_{ph}<1$ is obtained.

\subsection{Double occupancy}
We now turn to analyze the effects of the dot double occupancy on the
phonon distribution. We focus on the case of
$\lambda\lesssim 1$ and $A<1$ which, as we discussed above, favors
$F_{ph}<1$ when $\bar{U}$ is large. The choice $A<1$ is also
favorable for obtaining double occupancy, as tunnel--out rates are
suppressed. We will concentrate on the regime $\bar{U}>2\omega_{0}$
that is relevant in realistic situations with an intermediate
dot--phonon coupling strength. 
\begin{figure}[h]
\begin{center}
\includegraphics[width=12cm,keepaspectratio]{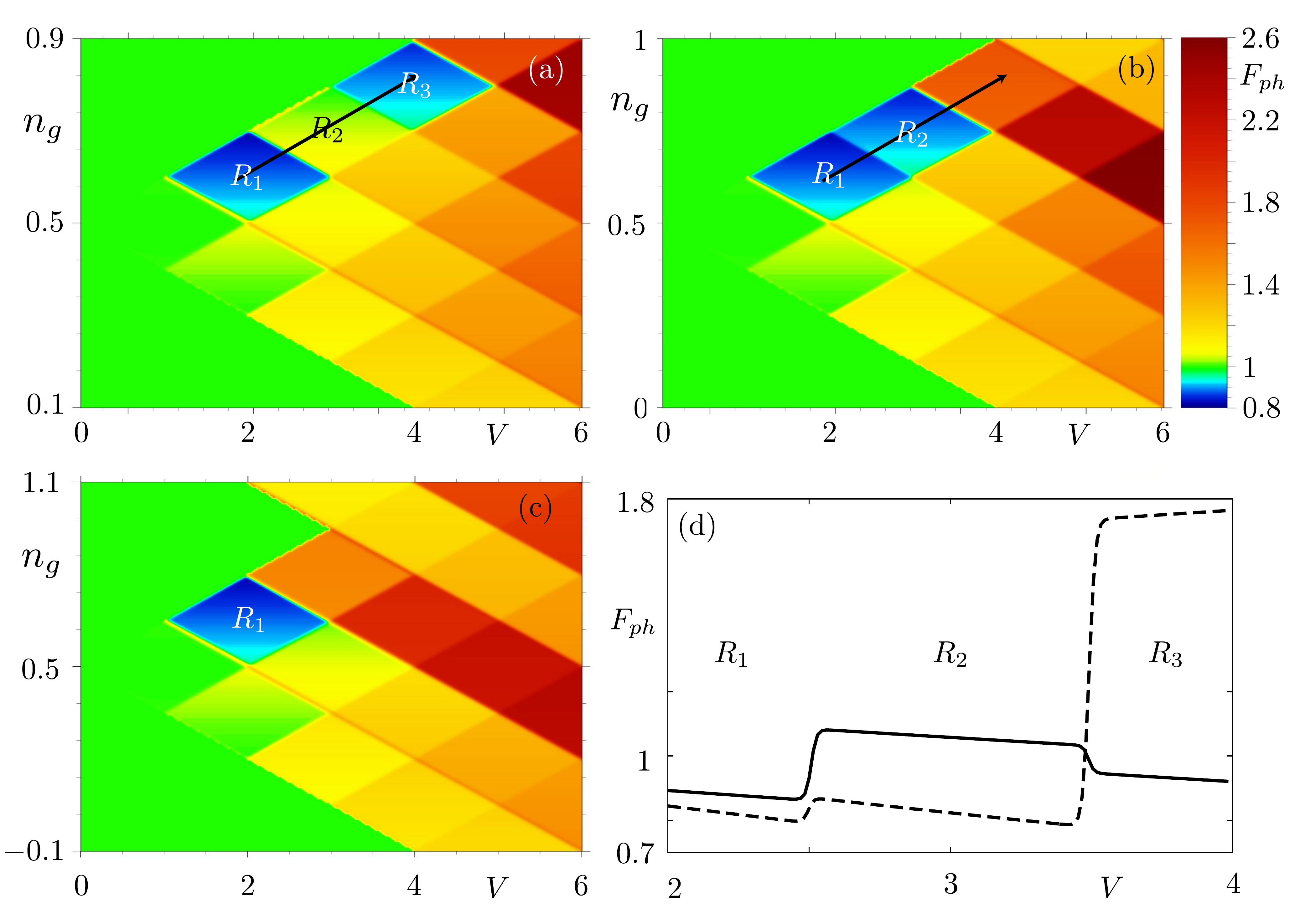}
\caption{(a--c) Color map of the phonon Fano factor $F_{ph}$ in the $V$
  (units $\omega_{0}/e$), $n_{g}$ plane for $A=0.1$, $\lambda=0.8$,
  $k_{B}T=0.01\ \omega_{0}$, $\Gamma_{L}=10^{-3}\omega_{0}$ and
  different values of $\bar{U}$: (a) $\bar{U}=5\ \omega_0$, (b)
  $\bar{U}=4\ \omega_0$, (c) $\bar{U}=3\ \omega_0$. (d) $F_{ph}$ as a
  function of $V$ (units $\omega_{0}/e$) calculated along the line
  $\bar{U}(1-2n_{g})+eV-\omega_{0}=0$ for $\bar{U}=5\ \omega_{0}$ (solid) and
  $\bar{U}=4\ \omega_{0}$ (dashed) (see the black arrows in panels
  (a), (b)).}
\label{fg:fig2}
\end{center}
\end{figure}
Figs.~\ref{fg:fig2}(a)--(c) show $F_{ph}$ for the same parameters as
in Fig.~\ref{fg:fig1}(a) but much smaller $\bar{U}$.  In addition to
region $R_1$, it is now $F_{ph}<1$ in region $R_{3}$ for $\bar{U}=5\
\omega_{0}$ and in region $R_{2}$ for $\bar{U}=4\ \omega_{0}$.  A plot
of $F_{ph}$ calculated along the diagonal line shown in
Figs.~\ref{fg:fig2}(a) and~\ref{fg:fig2}(b) is displayed in panel (c).
Note that the new regions where $F_{ph}<1$ shift at larger voltages
with increasing $\bar{U}$. This explains why the case
$\bar{U}\gg\omega_{0}$ is similar to the case $\bar{U}\to\infty$ near
the resonance $n=0\to 1$, where only the region $R_{1}$ displays
$F_{ph}<1$. The above behaviors represent special cases of the 
non--trivial general result:
for obtaining $F_{ph}<1$ in region $R_{p\geq 2}$ it is necessary to have
\begin{equation}
\label{eq:condfilt}
(p+1)\omega_{0}<\bar{U}<(p+3)\omega_0\,.
\end{equation}
In the following, we will explain the origin of Eq.~(\ref{eq:condfilt})
and show that it relates to specific transitions which involve the
state $n=2$.

Let us first of all determine the conditions under which double
occupancy can be achieved. In the absence of phonons ($\lambda=0$) the
onset of the transition $n=1 \to 2$ occurs for $V\geq
V_{0}=U(3-2n_{g})/e$. On the other hand, in the presence of phonons,
the state $|2,0\rangle$ can be occupied for smaller $V$: indeed, the
transitions $|1,l+q\rangle\to|2,q\rangle$ with $l\geq 1$ are active
for $V>V_l=[\bar{U}(3-2n_{g})-2l\omega_{0}]/e$, with
$V_{l}<V_{0}$. This phonon--mediated mechanism for obtaining a double
occupancy is similar to that described by Shen {\em et
al.}~\cite{shen} for the occupation of the LUMO in a molecular quantum
dot. The condition for having the transition
$|1,l+q\rangle\to|2,q\rangle$ open on the left barrier (forward
transition, relevant for $V>0$) is
\begin{equation}
\label{eq:deltaE3}
\bar{U}(1-2n_{g})-2\omega_{0}l+2\bar{U}-eV\leq 0\, .
\end{equation}
In view of voltages constraints (\ref{eq:centerRp}) for region
$R_{p}$, the above transitions open there for
$\bar{U}<(p+l+1)\omega_{0}$. We will now show that double occupancy
can lead to a selective phonon population with
$\bar{P}_{0}\approx\bar{P}_{1}\gg\bar{P}_{l\geq 2}$ in the regions
$R_{p\geq 2}$. For $\bar{U}<(p+3)\omega_{0}$ the transitions
\begin{equation}
\label{eq:depop1}
|1,l+q\rangle\to|2,q\rangle\quad\quad l\geq 2
\end{equation}
are active within $R_{p}$ and decrease the occupation probability of
the excited phonon bath states with $l\geq 2$. They come in addition
to the usual transitions
\begin{equation}
\label{eq:depop2}
|1,l\rangle\to|0,l'\rangle\quad\quad l'<l
\end{equation}
which also depopulate the excited phonon states and were already
present when $\bar{U}\gg\omega_{0}$. Note that these latter
transitions alone {\em are not} sufficient to give rise to a selective
phonon population: indeed, broad phonon distributions are obtained for
$\lambda<1$ in $R_{p\geq 2}$ as already discussed in
Sec.~\ref{sec:nodouble}. Now, since for $A<1$ and $V>0$ tunnel--in
events are faster than tunnel--out ones, the new
transitions~(\ref{eq:depop1}), are much more effective than
transitions~(\ref{eq:depop2}) in depopulating states with $l\geq
2$. With these new fast channels it is now possible to obtain lower
occupations of all the excited phonon bath states up to $l=2$,
eventually leading to a selective phonon population. 

\noindent This explains the right hand side of the
inequality~(\ref{eq:condfilt}).  To obtain the left hand side
of~(\ref{eq:condfilt}) it is sufficient to recognize that if
$\bar{U}<(p+1)\omega_{0}$ the additional depopulation transitions
$|1,l+q\rangle\to|2,q\rangle$ with $l\geq 1$ are open in all $R_{p}$.
They also deplete the states with $l=1$, eventually leading to too a
narrow distribution with $\bar{P}_{0}\approx 1$ and $F_{ph}>1$.

\noindent Therefore, tuning $\bar{U}$  it
is possible to trim the {\em shape} of phonon population from broad,
to selective, up to a very narrow distribution.
\begin{figure}[h]
\begin{center}
\includegraphics[width=6cm,keepaspectratio]{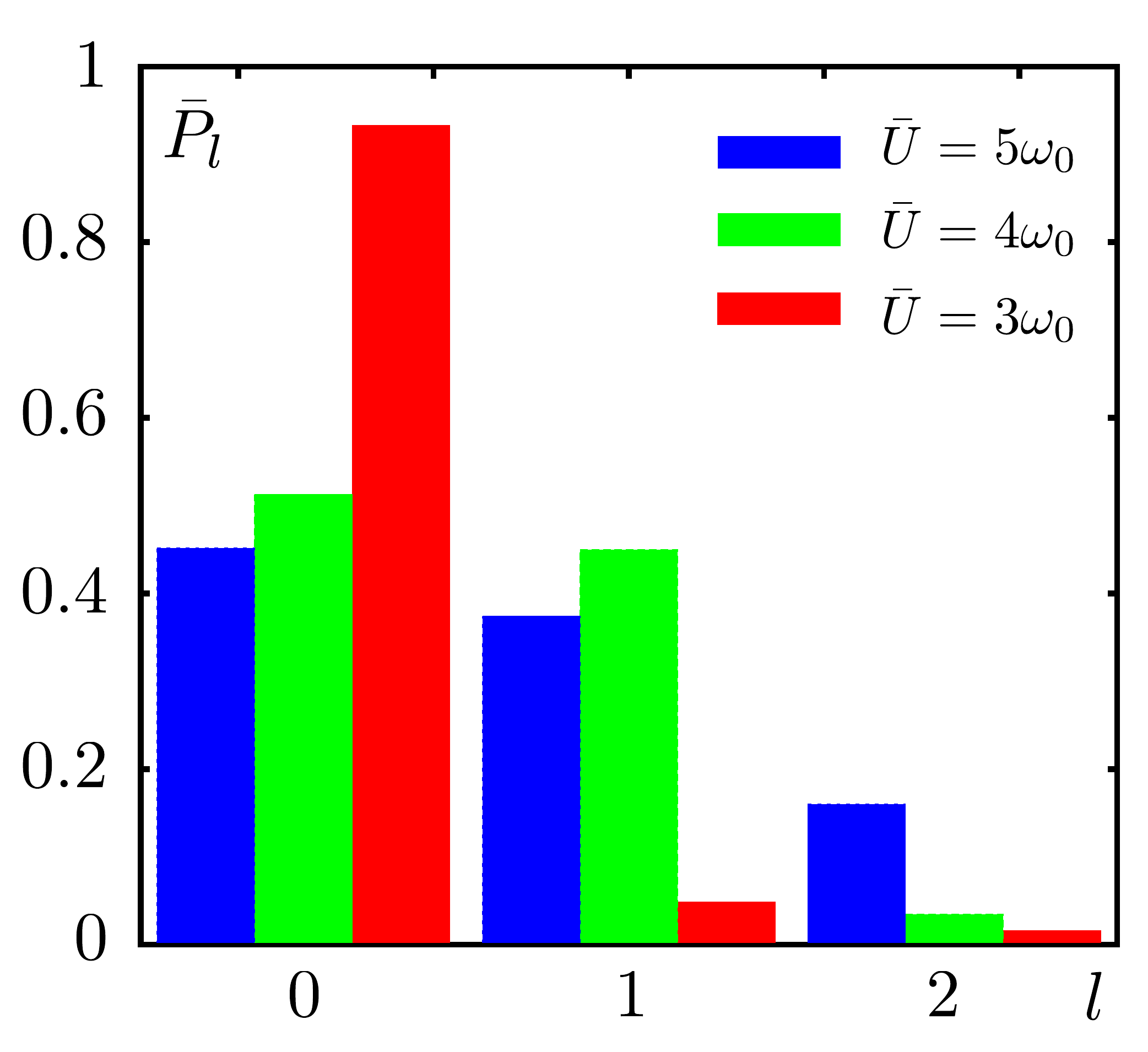}
\caption{Phonon distribution $\bar{P}_{l}$ as a function of $l$
  calculated in the center of region $R_{2}$ for different values of
  $\bar{U}$, all other parameters as in Fig.~\ref{fg:fig1}(a). For
  $\bar{U}=5\omega_{0},4\omega_{0},3\omega_{0}$ it is $F_{ph}\approx 1.1,0.8,1.5$, respectively.}
\label{fg:fig3}
\end{center}
\end{figure} 
Figure~\ref{fg:fig3} shows $\bar{P}_l$ as a function of $l$,
calculated in the center of $R_{2}$ for different $\bar{U}$, all other
parameters are the same as in Fig.~\ref{fg:fig1}(a). For
$\bar{U}=5\omega_{0}$, the transitions $|1,2+q\rangle\to|2,q\rangle$
are closed and a {\em broad} phonon distribution (blue bars) with
$F_{ph}>1$ is obtained. When $\bar{U}=4\ \omega_{0}$, the above
transitions are open and the channel $|1,2\rangle\to|2,0\rangle$
creates the selective population displayed by the green bars. Further
reducing $\bar{U}=3\ \omega_{0}$, also the transitions
$|1,1\rangle\to|2,0\rangle$ opens up and the resulting narrow
distribution displays super--Poissonian character.

\section{Damping effects and the validity of the RWA}
\label{sec:stability}
In Section~\ref{sec:rwa} we have discussed the out of equilibrium
localized bath arising in the transport regime, concentrating
particularly on the occurrence of sub--Poissonian phonon
distributions. The results were derived assuming that the localized
phonon mode has an infinite life time in the absence of tunneling. In
addition, up to now we have always considered the case of a fast
motion of the oscillator in comparison to the transport time scales
and adopted the RWA. In this section we study the robustness of the
results, when the above conditions are relaxed.

\subsection{Towards a relaxed single mode bath}
\label{sec:relax}
In order to damp the single mode bath towards an equilibrium
distribution, we couple it to a conventional dissipative environment,
represented by a set of harmonic oscillators
\begin{equation}
\label{eq:bath3}
H_{b}^{(3)}=\sum_{j}\omega_{j}b^{\dagger}_{j}b_{j} 
\end{equation}
with a linear coupling
\begin{equation}
\label{eq:coup}
H_{bb}^{(13)}=(b+b^{\dagger})\sum_{j}\chi_{j}\omega_{j}(b_{j}+b^{\dagger}_{j})
\end{equation}
and bath spectral density
$J(\omega)=2\pi\sum_{j}\omega^{2}_{j}\chi_{j}^{2}\delta(\omega-\omega_{j})$.
In the presence of the additional bath~(\ref{eq:bath3}), it is still
possible to diagonalize exactly $H_{\lambda}$ by means of a canonical
transformation which also involves the operators $b_{j}$,
$b_{j}^{\dagger}$~\cite{braig,haupt}. Following a derivation analogous
to that discussed in Sec.~\ref{sec:master} and treating both
Eq.~(\ref{eq:tunnel2}) and Eq.~(\ref{eq:coup}) to the lowest
perturbative order, we obtain the following stationary rate equations
in the RWA~\cite{shen,haupt}
\begin{eqnarray}
\label{eq:master_rel}
&&-z_{n}\bar{P}_{nq}\sum_{k=\pm 1\atop 0\leq n+k\leq
  2}\sum_{p=0}^{\infty}\Gamma_{qp}^{n,n+k}+\sum_{k=\pm 1\atop 0\leq
  n+k\leq
  2}\sum_{p=0}^{\infty}z_{n+k}\bar{P}_{n+k,p}\Gamma^{n+k,n}_{pq}\nonumber\\ &&+\sum_{p=0}^{\infty}\bar{P}_{np}
\Gamma^{rel}_{pq}-\sum_{p=0}^{\infty}\bar{P}_{nq}\Gamma^{rel}_{qp}=0\, . 
\end{eqnarray}
The relaxation rates
\begin{equation}
\Gamma^{rel}_{q,q-1}=e^{\beta\omega_0}\Gamma^{rel}_{q-1,q}=qw\Gamma_{L}{\left(1-e^{-\beta\omega_0}\right)}^{-1}
\end{equation}
induce transitions $|n,q\rangle\to|n,q\pm 1\rangle$ and drive the
single mode bath towards thermal equilibrium, while the tunneling
rates $\Gamma_{pq}^{nn'}$ are those described in
Sec.~\ref{sec:rwa}. Here, $w=J(\omega_0)/\Gamma_L$ parametrizes the
relaxation strength. For $w\to\infty$, phonons are completely relaxed
and $\bar{P}_{l}\to
\bar{P}_{l}^{(th)}$, characterized by a super--Poissonian Fano factor. 
\begin{figure}[h]
\begin{center}
\includegraphics[width=15cm,keepaspectratio]{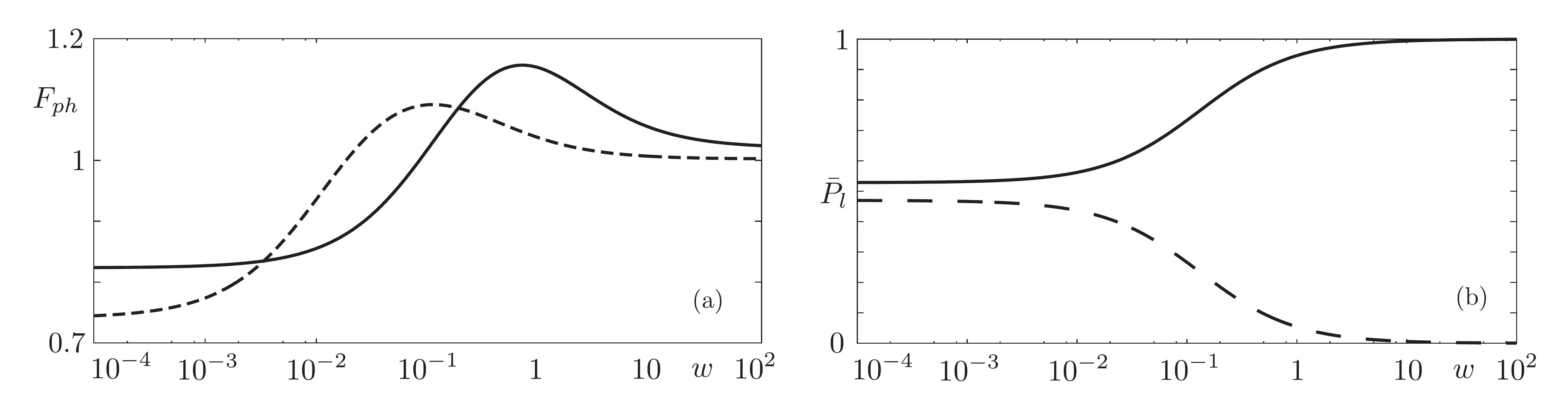}
\caption{(a) Phonon Fano factor $F_{ph}$ for $A=0.1$ (solid) and
  $A=0.01$ (dashed) as a function of the relaxation strength $w$ (log
  scale) for $\bar{U}=4\ \omega_{0}$, $eV=3\ \omega_{0}$ and
  $n_{g}=0.75$ (the center of $R_{2}$). All other parameters as in
  Figs.~\ref{fg:fig2}(a)--(c). (b) Phonon populations $\bar{P}_{0}$ (solid),
  $\bar{P}_{1}$ (dashed) as a function of $w$ (log scale) for the same parameters as in (a).}
\label{fg:fig4}
\end{center}
\end{figure}

\noindent The effects of relaxation on the phonon distribution are
illustrated in region $R_{2}$ for $\bar{U}=4\ \omega_{0}$. All other
regions display qualitatively similar behavior. Fig.~\ref{fg:fig4}(a)
shows $F_{ph}$ as a function of $w$, calculated in the center of
$R_{2}$. As $w$ increases,
$F_{ph}$ crosses from $F_{ph}<1$ in the unequilibrated regime to a
super--Poissonian maximum, before reaching the thermal value which,
for $T\ll\omega_{0}/k_{B}$, is only slightly above 1. The effects of
relaxation become relevant when the typical life time of the excited
phonon bath $\tau_{rel}=J(\omega_{0})^{-1}$ is comparable with
the {\em slowest} typical charge dwell time
$\tau_{tr}=(\mathrm{min}\{1,A\}\Gamma_L e^{-\lambda})^{-1}$. This
explains also the shift of the maximum towards smaller $w$ for smaller
asymmetries observed in Fig.~\ref{fg:fig4}(a). In
Fig.~\ref{fg:fig4}(b), one can see that for increasing relaxation, the selective phonon population is destroyed with a tendency
towards an almost full occupation of the $l=0$ state, in accordance
with a low temperature $\bar{P}_{l}^{(th)}$ (see Eq.~(\ref{eq:thermal}).
\begin{figure}
\begin{center}
\includegraphics[width=15cm,keepaspectratio]{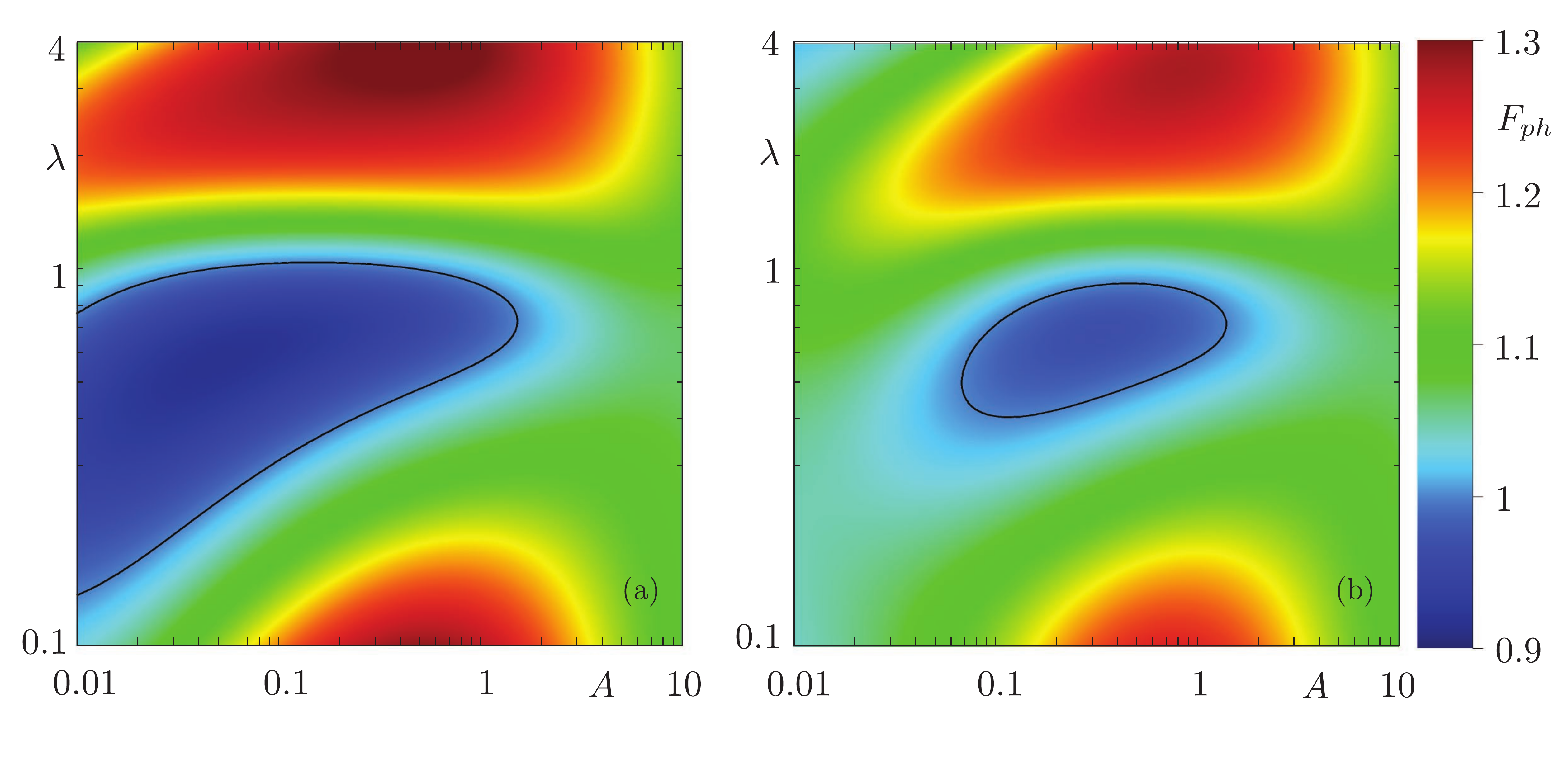}
\caption{(a) Color map
  of $F_{ph}$ as a function of $A$, $\lambda$ (both in log scale), for
  $eV=2\ \omega_{0}$, $n_{g}=0.625$ (region $R_1$), $\bar{U}=4\ \omega_{0}$ and
  $w=0.01$, other parameters as in Figs.~\ref{fg:fig2}(a)--(c). The black
  contour line is set at $F_{ph}=1$. (b) Same as panel (a) but with
  $w=0.1$.}
\label{fg:fig4_2}
\end{center}
\end{figure} 

We now analyze the robustness of the sub--Poissonian phonon bath
against $w$ varying the parameters $A$,
$\lambda$. Figures~\ref{fg:fig4_2}(a) and \ref{fg:fig4_2}(b) show a color map of $F_{ph}$ as
a function of $A$ and $\lambda$ in the middle of the region $R_{1}$ for two different values of the
relaxation strength. 
Superimposed to the color map, the black contour line
signals $F_{ph}=1$. This line corresponds to the ``stability
boundary'' of the sub--Poissonian phonon distribution with respect to
asymmetry and coupling strength. Increasing relaxation, Fig.~\ref{fg:fig4_2}(b), 
the region of $A$ and  $\lambda$ where  $F_{ph}<1$ shrinks, the regime $A\ll 1$ is more
strongly affected, in accordance with the discussion above, while the sub-Poissonian distribution is still stable  
for not too small $A<1$ and $\lambda\approx 1$. Qualitatively analogous results are observed in every region
$R_{p}$.
\subsection{Validity of the RWA}
\label{sec:coherence}
When the dynamics of the single mode bath is not much faster than the
typical electron tunneling rate the RWA is no longer justified and the fully
coherent GME in Eq.~(\ref{eq:gme_schro}) must be solved. In this case,
also the non diagonal elements of the density matrix will be different
from zero in the stationary regime. Even more important, while in
the RWA the diagonal elements $\bar{\rho}_{ll}^{n}$ of the density
matrix are {\em decoupled} from the off--diagonal ones, in  general
this is no longer the case. Thus the coherences
$\bar{\rho}_{l,l'\neq l}$ may influence the phonon distribution 
$\bar{P}_{nl}$. In this last part, we will present preliminary results 
showing this effect.
For simplicity, we consider $\bar{U}\gg\omega_{0}$, without relaxation $w=0$.
\begin{figure}[h]
\begin{center}
\includegraphics[width=15cm,keepaspectratio]{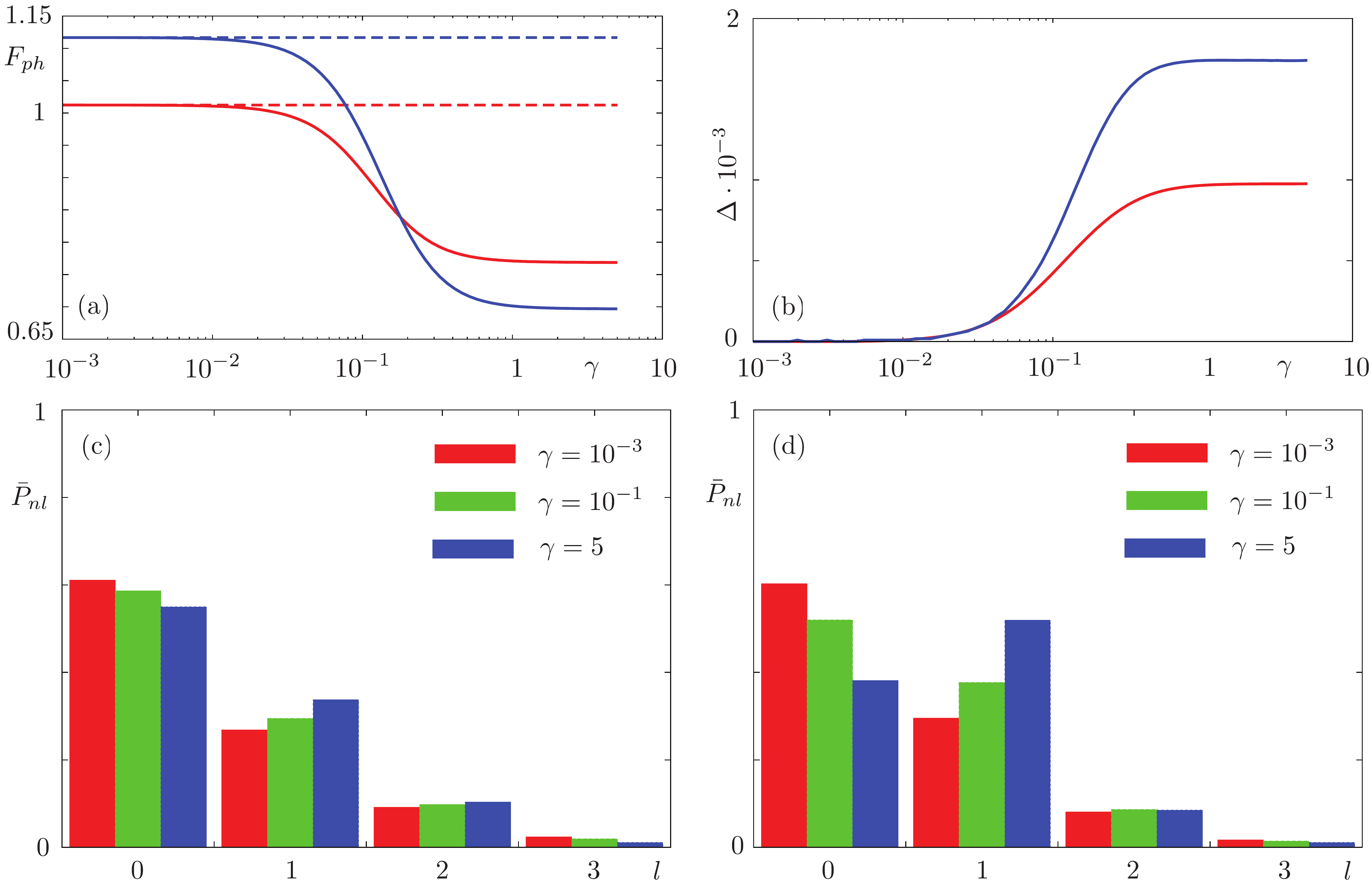}
\caption{(a) Solid: phonon Fano factor $F_{ph}$ obtained from the
  solution of Eq.~(\ref{eq:gme_schro}) as a function of
  $\gamma=\Gamma_{L}/\omega_{0}$ for $A=0.01$, $\lambda=0.1$, $T=0.1\
  \omega_{0}/k_{B}$, $\bar{U}=20\ \omega_{0}$, and $eV=2\ \omega_{0}$
  with $n_{g}=0.525$ (red, in region $R_{1}$) or $eV=3\ \omega_{0}$
  with $n_{g}=0.5$ (blue, in region $S$). Dashed: $F_{ph}$ obtained
  from the RWA. (b) Relative difference $\Delta$ between $F_{ph}$ and
  $F_{ph}^{0}$ (see text for definition), as a function of $\gamma$
  for the same parameters as panel (a). (c) Diagonal elements
  $\bar{P}_{l}$ of the solution of Eq.~\ref{eq:gme_schro} as a
  function of $l$ calculated for $eV=2\ \omega_{0}$ and $n_{g}=0.525$
  and different values of $\gamma$. (d) Same as (c) but for $eV=3\
  \omega_{0}$ and $n_{g}=0.5$.}
\label{fg:fig5}
\end{center}
\end{figure} 

\noindent Fig.~\ref{fg:fig5}(a) shows $F_{ph}$ obtained from the solutions of
the full GME in Eq.~(\ref{eq:gme_schro}) as a function of
$\gamma=\Gamma_{L}/\omega_{0}$, where we keep $\omega_{0}$ fixed (note
that the calculation of $F_{ph}$ via
Eqs.~(\ref{eq:avgl}) and~(\ref{eq:avgl2}) contains also averages of
non--diagonal operators). The red curve corresponds to $F_{ph}$ in
the center of region $R_{1}$, while the blue one gives $F_{ph}$ in region
$S$. The dashed lines correspond to the RWA solution. In both cases,
it is clear that deviations from the RWA (dashed line) occur for
$\gamma>0.05$. For increasing values of $\gamma$, $F_{ph}$ saturates
into a new regime, characterized by sharply reduced phonon
fluctuations.

In order to assess the relevance of the off--diagonal density matrix
elements, we have evaluated the ratio
$\Delta =|F_{ph}-F^{0}_{ph}|/F_{ph}$, where $F_{ph}^{0}$ is
calculated from the GME solution {\em neglecting the coherences}
$\bar{\rho}_{l,l'\neq l}^{n}$. We show $\Delta$ in
Fig.~\ref{fg:fig5}(b): clearly, the impact of the off--diagonal terms
on $F_{ph}$ is very small, with $\Delta<2\cdot 10^{-3}$. This
fact is supported also by direct inspection of the largest
$\bar{\rho}_{l,l'\neq l}^{n}$, whose absolute values are three to four
orders of magnitude smaller than the leading diagonal terms. The
origin of the reduction of $F_{ph}$ in Fig.~\ref{fg:fig5}a therefore lies in a modification
of $\bar{P}_{nl}$, {\em mediated} by the off--diagonal terms of the
density matrix.

\noindent In Fig.~\ref{fg:fig5} we show the population of the phonon bath $\bar{P}_{l}$ 
obtained in $R_{1}$, panel (c), and in $S$, panel (d). The
changes mentioned above can be clearly seen: in $R_{1}$, for
increasing $\gamma$ the phonon population gets broader.
In $S$ an
{\em inversion} of the population for the levels $l=0$ and $l=1$ is
present, with an even stronger sub--Poissonian $F_{ph}$.

The preliminary results show that there are regimes 
where dynamical coherent effects are relevant for the single phonon 
distribution. Much in the same way as for $\bar{U}$,
phonon distributions can be dramatically shaped by tuning $\gamma$. Strictly speaking, the results for $\gamma>0.1$
and $k_{B}T=0.1\omega_0$ imply $\Gamma_{L}>k_{B}T$ and therefore fall beyond the
limits of the sequential tunneling regime for the calculation
presented above. In this case we can expect that the additional
coherent dot dynamics could lead to even larger deviations with
respect to the RWA solutions.
\section{Conclusions}
In this paper we have analyzed the out of equilibrium properties of an
unconventional single mode phonon bath, coupled to a quantum dot, in
terms of its phonon distribution and the respective Fano factor. We
have described the combined evolution of the dot and the bath degrees
of freedom by means of a GME. In the regime of validity of the RWA,
corresponding to a {\em fast} bath motion, we have shown that the out
of equilibrium phonon bath markedly differs from a thermal one. The
transport properties of the dot in the presence of a thermal or an
unequilibrated single bath mode are qualitatively different. In the case
of a single occupancy of the quantum dot, we have identified the
transport regimes and the range of asymmetries $A$ and dot--bath
coupling strengths $\lambda$ for which a sub--Poissonian phonon
distribution appears. We have interpreted such distribution in terms
of a peculiar, {\em selective} population of the bath modes.  Double
occupancy of the dot may act as a filter on the phonon bath
distribution, eventually turning $F_{ph}<1$ in regions where a
super--Poissonian distribution would be attained in the case of a
single dot occupancy. The crossover from an unequilibrated phonon bath
to a thermal regime has been investigated by coupling the single bath
mode to an external dissipative environment. For increasing damping
strength, the sub--Poissonian phonon bath is destroyed and the latter
tends to a thermal one. The crossover from a sub-- to
super--Poissonian bath has been analyzed as a function of $A$ and
$\lambda$. We have found that $F_{ph}<1$ survives up to relatively
strong damping when $A\approx 1$ and $\lambda\approx 1$. Finally, we
have shown that in the case of $\omega_{0}$ not much larger than
the tunneling rates, the RWA is no longer valid
and the solutions of the full GME display strong modifications when
the coherent dynamics of the bath is fully taken into account.  In
particular, we have shown that although off--diagonal density matrix
elements remain small in the coherent regime, their coupling to the
diagonal elements gives rise to significant modifications of the
latter. These results show that by suitably tuning the quantum dot
properties, it is possible to obtain a variety of unconventional
single phonon baths characterized by a wide range of phonon
distributions. A detailed analysis of the coherent regime and the
possible extension to out of equilibrium baths with two or few more
phonon modes is deferred to future work.
\section{Acknowledgements}
The authors acknowledge stimulating discussions with F. Haupt,
M. Merlo and A. Braggio. Financial support by the EU via contract
no. MCRTN-CT2003-504574 and by the Italian MIUR via PRIN05 is
gratefully acknowledged.

\end{document}